\documentclass[sigconf]{acmart}
\AtBeginDocument{%
  \providecommand\BibTeX{{%
    \normalfont B\kern-0.5em{\scshape i\kern-0.25em b}\kern-0.8em\TeX}}}

\usepackage{url}
\usepackage{pdfpages}

\acmYear{2021}\copyrightyear{2021}
\setcopyright{acmcopyright}
\acmConference[GoodIT '21]{Conference on Information Technology for Social Good}{September 9--11, 2021}{Roma, Italy}
\acmBooktitle{Conference on Information Technology for Social Good (GoodIT '21), September 9--11, 2021, Roma, Italy}
\acmPrice{15.00}
\acmDOI{10.1145/3462203.3475910}
\acmISBN{978-1-4503-8478-0/21/09}

\begin{document}
\includepdf[pages={1}]{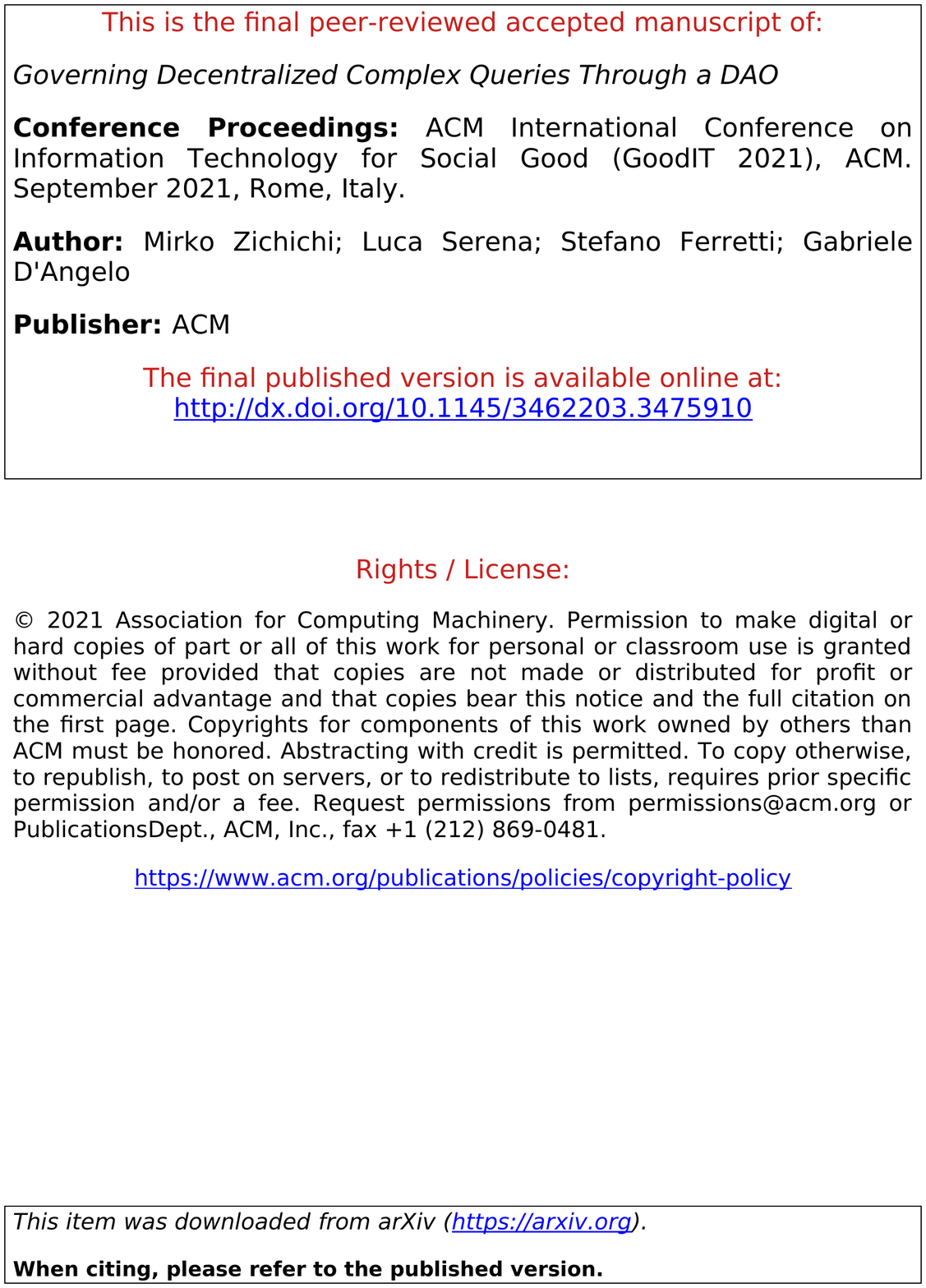}

\title{Governing Decentralized Complex Queries Through a DAO}

\author{Mirko Zichichi}
\thanks{This work has received funding from the EU H2020 research and innovation programme under the MSCA ITN European Joint Doctorate grant agreement No 814177 Law, Science and Technology Joint Doctorate - Rights of Internet of Everything.}
\affiliation{%
  \institution{Ontology Engineering Group}
  \city{Universidad Politécnica de Madrid}
  \country{Spain}}
\email{mirko.zichichi@upm.es}

\author{Luca Serena}
\affiliation{%
  \institution{CIRI ICT}
  \city{University of Bologna}
  \country{Italy}}
\email{luca.serena2@unibo.it}

\author{Stefano Ferretti}
\affiliation{%
 \institution{Department of Pure and Applied Sciences}
 \city{University of Urbino ``Carlo Bo"}
 \country{Italy}}
\email{stefano.ferretti@uniurb.it}

\author{Gabriele D'Angelo}
\affiliation{%
  \institution{Department of Computer Science and Engineering}
  \city{University of Bologna}
  \country{Italy}}
\email{g.dangelo@unibo.it}

\renewcommand{\shortauthors}{Zichichi et al.}

\begin{abstract}
Recently, a new generation of P2P systems capable of addressing data integrity and authenticity has emerged for the development of new applications for a "more" decentralized Internet, i.e., Distributed Ledger Technologies (DLT) and Decentralized File Systems (DFS).
However, these technologies still have some unanswered issues, mostly related to data lookup and discovery.
In this paper, first, we propose a Distributed Hash Table (DHT) system that efficiently manages decentralized keyword-based queries executed on data stored in DFS. Through a hypercube logical layout, queries are efficiently routed among the network, where each node is responsible for a specific keywords set and the related contents. 
Second, we provide a framework for the governance of the above network, based on a Decentralized Autonomous Organization (DAO) implementation.  We show how the use of smart contracts enables organizational decision making and rewards for nodes that have actively contributed to the DHT.
Finally, we provide experimental validation of an implementation of our proposal, where the execution of the same protocol for different logical nodes of the hypercube allows us to evaluate the efficiency of communication within the network.
\end{abstract}

\ccsdesc[500]{Computer systems organization~Distributed architectures}
\ccsdesc[500]{Information systems~Distributed storage}

\keywords{Distributed Ledger Technology, Decentralized File Storage, Distributed Hash Table, Keyword Search, Smart Contracts}

\maketitle

\section{Introduction}
The digitalization process, that has been ongoing over the last decades, has seen data management and delivery become a crucial issue.
In order to cope with the increasingly higher number of contents that is demanded through the Web, multiple solutions for an efficient use of Internet have been designed. 
In particular, thanks to the decentralization of content storage and delivery, it is possible to avoid single points of failures, reduce the workload at data centers and to allow a distribution of data that is closer to the original source. Decentralization also fosters the creation of open systems, where participants can freely join the system and contribute to its functioning.

Recently, Distributed Ledger Technologies (DLTs) and Decentralized File Systems (DFS) 
have emerged as Peer-to-Peer (P2P) technologies capable of offering interesting features related to data validation and trustfulness~\cite{zichichi2020framework}. DLTs have gained popularity with the advent of the cryptocurrencies, which allow users to trade crypto-assets without any central entity being involved, ensuring transparency and data integrity. 
Besides the financial use case, DLTs, and DFS in particular, provide the features of data integrity, authenticity, confidentiality and auditability, used to build novel applications for a ``more'' decentralized Internet \cite{belotti2019vademecum}.

InterPlanetary File System (IPFS) is one of the most used DFS protocols
where files and data are replicated globally on hundreds of nodes in the network
\cite{benet2014ipfs}. 
Furthermore, DFS and in general other P2P-based technologies might also have a prominent role against censorship, since shutting down a server will not prevent contents from being available on Internet. An example occurred when Turkey denied the access to the Turkish Wikipedia in 2017, with IPFS being able to guarantee the access through mirroring \cite{santos2019dclaims}.

One of the aspects that remains still open with respect to these novel technologies, is concerned with the data discovery and lookup. 
Specifically, data can only be accessed by knowing the respective identifier or location and cannot be searched based on its content. 
Put in other words, these systems lack a viable (decentralized) data management scheme that enables ``complex'' queries on top of them.

In this paper, we propose a decentralized system to efficiently manage  keyword-based queries to the contents stored in DFS. We make use of a Distributed Hash Table (DHT) structured as a hypercube in order to provide the service of keyword-based queries over IPFS files. The hypercube is a logical layout where there are $2^r$ nodes, each one labelled with an $r$ bits ID and connected to the $r$ nodes whose ID differs by only one bit. Each node is responsible for a specific keywords set, derived from their ID. The hypercube structure allows to optimize the routing of the queries, by reducing the number of hops needed to locate contents. Moreover, the second main contribution of this paper is the creation of a framework for the organization of nodes operators that host and share information, with the aim of improving the scalability and the decentralization of the system. In fact, usually applications built upon P2P networks are supported by nodes that have no particular incentive to keep them operational, but are only interested in their use, e.g.~BitTorrent. In our work, we focus on the case where nodes are interested in keeping the network operational and healthy, and, in general, where node operators act in the context of a sharing economy, e.g.~in the same way as Wikipedia editors are interested in contributing to the free encyclopedia \cite{hamari2016sharing}.
It has been argued that DLTs can ``crystallize'' the dynamics of a model of socio-economic production in which large numbers of people work cooperatively, i.e.~commons-based peer production \cite{pazaitis2017blockchain}. P2P networks have this ``cooperative vein'' intrinsically built into their structure. Therefore, we leverage DLTs to build a Decentralized Autonomous Organization (DAO)~\cite{jentzsch2016decentralized} around the network of node operators. 
We envision an approach that is based on the creation of a DAO for those actors who have actively contributed to the functioning of the system, with smart contracts involved in managing rewards and organizational decisions. We then propose different use cases where this network can take form and we provide a possible framework for its governance.

Finally, we report an experimental validation of an implementation of our proposal.
In particular, we provide results showing how the size of the hypercube and the number of objects stored in the DHT affect the search procedures. Moreover, we also evaluate the smart contracts, implemented to be executed on the Ethereum blockchain, in terms of operations execution cost.

This paper is structured as follows. Section \ref{sec:back} provides the background on the technologies used. Section \ref{sec:keyw} presents a description of the hypercube DHT structure, while in Section \ref{sec:dao} we argue on how such a system can be governed through a DAO. In Section \ref{sec:valid}, the system experimental validation is reported, finally, Section \ref{sec:concl} provides the concluding remarks.

\section{Background and Related Work}\label{sec:back}

\subsection{Distributed Hash Tables (DHTs)}
A Distributed Hash Table (DHT) is a decentralized system for the distributed storage of contents. The rationale of this approach is to store the information in the various nodes of the system, providing a routing mechanism to easily get which node owns a certain resource \cite{joung2007keyword}. Each local view of the DHT nodes will look like a traditional hash table, with a mapping from a key (i.e.~the univocal representation of an item) to values (i.e.~addresses of the peers owning such a resource).
The association of objects to DHT nodes is obtained through the use of a hash function, a one-way function which maps any item into a binary sequence of $n$ bits. The idea is to distribute the storage workload among the DHT nodes according to the key (i.e.~the $n$ bit string obtained after having applied the hash function) of the objects. Each DHT is identified itself through an $n$ bit ID, which lies in the same ID space used to identify contents. Then, based on its ID, each node is in charge of maintaining information on those contents that are in a specific ID space interval. Lookup of a content $x$ thus becomes looking for the node in the DHT that manages a subset of the ID space that contains $x$ \cite{d2017highly}.

\subsection{Distributed Ledger Technology (DLT)}
A DLT is a P2P system where the participants maintain a copy of the ledger, and there is a consensus mechanism that allows all the nodes to have the same view on the stored information. Data written on the ledger are trustworthy, because DLT protocols ensure their integrity, immutability and authenticity. 
There are multiple DLT implementations, differing mostly for their structure (e.g.~blockchain, DAG)
and for the features they provide, such as smart contracts.

\subsubsection{Smart Contracts and Decentralized Autonomous Organizations}
Smart contracts are programs whose execution is performed in a distributed way. In Ethereum~\cite{buterin2013ethereum}, all the participants receive the same inputs and perform a computation on the basis of a smart contract code that leads to the same outputs. Each process is thus completely traced and permanently stored on the blockchain.
Smart contracts can be used to automatize and supervise the exchange of digital or physical assets, to create tokens, i.e. the representation of physical assets or utilities, and to allow the management of a Decentralized Autonomous Organization (DAO)~\cite{jentzsch2016decentralized}.
In order to enable a decentralized management of a DAO, smart contracts implement transactions, currency flows, rules and rights within the organization. DAO members can make proposals for the management of the organization and also discuss and vote those through transparent mechanisms. Members can also interact through smart contracts and tokens can be sent or received. Usually, tokens grant their holder a certain set of rights within the DAO.

\subsection{Decentralized File Storage (DFS)} 
Decentralized File Storages (DFS) offer an alternative to the traditional client-server models, i.e. where a domain name is provided and is then resolved to an IP address.
In Content Based Addressing items are directly queried through the network rather than establishing a connection with a server. In order to know which node in the network own the requested contents, it is possibile to rely on a DHT system that is in charge of mapping the items with the addresses of the peers owning such data.
DFS follow this approach and offer higher data availability and resilience thanks to data replication.

\subsubsection{IPFS and File Search}
The InterPlanetary File System (IPFS) is a DFS and a protocol thought for distributed environments with a focus on data resilience~\cite{guidi2021data}.
The P2P network that runs the IPFS protocol, stores and shares files in the form of IPFS objects that are identified by a CID (Content IDentifier). This CID consists in the digest produced when a hash function is applied to a file and it is used to retrieve the referenced IPFS object. However, it provides no means of searching for a file without owning it, since its hash is required. To overcome this limitation a generic search engine has been developed, namely ``ipfs-search'' \cite{ipfssearch2021}. This solution is rather centralised and does not escape the problem of concentration similar to the conventional web.
In response to this, a decentralized solution called Siva \cite{khudhur2019siva} has been proposed. An inverted index of keywords is built for the published contents on IPFS and users can search through it, however Siva is proposed as an enhancement of the IPFS public network DHT and does not feature any optimization for a keyword storage structure apart from the use of caching.
In terms of aim, a system, similar to the one presented in this paper, is The Graph, a ``Decentralized Query Protocol'' \cite{thegraph2020protocol}. The Graph network consists of a system built upon Ethereum and IPFS, that allows to query data stored in these two technologies. 
The organization of the network is similar to what is referred as DAO, however their method for storing indexes is different from our proposal.

\section{Multiple Keywords Search}\label{sec:keyw}
The hypercube geometric form has been leveraged by Joung et al. \cite{joung2007keyword} to organise the topological structure of a DHT network, by using keywords. Such DHT can be exploited to perform multiple keyword based queries. Let $K \subseteq W$ represent a keywords set in a keyword space $W$. Such set $K$ can be used to perform a query to search for data contents characterized by keywords contained in $K$ (e.g.~as metadata). In particular, let $O$ be a set of data objects referenced in the DHT, these are distributed among all the network nodes based on the keywords they have been associated with, i.e. all the objects $o \in O$ mapped to a keywords set $K_o \subseteq W$ are maintained by the node responsible for such a keywords set $K_o$. We then consider $O$ as the set of all the CIDs in IPFS and we use the DHT to map keywords to IPFS Objects.

\subsection{Hypercube}
The DHT takes the form of a $r$-dimensional hypercube $H_r(V,E)$, where $V$ is set of vertices representing logical network nodes and $E$ is a set of edges that connect pairs of neighbor nodes. 
The ID of a logical node is given by an $r$-bit string associated to the keywords set the node is responsible for. 
Each bit position refers to a specific keyword. Thus, let assume that a given keywords set $K$ contains a given keyword $k$, which is assigned to the $i$-th bit of the string by a function. Then, the in the $r$-bit string representation of $K$ the $i$-th bit will be set to $1$. More formally, 
keywords are given in input to an uniform hash function $h : W \rightarrow \{0, 1, \hdots,r-1\}$ that returns positions to set to 1 in $r$-bit string, i.e. $\textit{one}(u) = \{ h(k) \mid k \in K \}$. 

We leverage such a protocol and make use of these $r$-bit strings to identify logical nodes in our system, e.g. for $r=4$, a node that has ID $0100$ handles all those the objects $o \in O$ associated to keywords sets whose the \textit{one} function returns $0100$.
In the DHT, connections between nodes, i.e. edges, are created among those nodes whose IDs differ of only one bit, e.g. $1011$ and $1010$. This link creation method builds an hypercube structured graph.

\subsection{Keywords Queries}
In our system, the discovery of contents in IPFS is based on the lookup of multiple keywords stored in the DHT. In particular each logical node will locally store an index table where the CIDs of all the IPFS Objects associated to the keywords set it is responsible for are stored. 
Given a query for a keywords set $K$, the associated $r$-bit string is used to reach the responsible logical node through a routing mechanism based on the hypercube form. 

Take, for instance, a keyword space $W$ made of $6$ keywords, $W=\{\textit{``Temperature'', ``PoI'', ``Wikipedia'', ``Bologna'', ``Urbino'', ``Rome''}\}$.\\
Let consider a query with a keywords set $K=\{$\textit{``Wikipedia, Rome''}$\}$ and assume that the $r$-bit query string associated to $K$ will be $001001$. Thus, to answer the query the node $u$ with ID $001001$ must be contacted. Starting from a node $v \in V$ in the hypercube, the query process consists in passing the request from $v$ to one of its neighbours that is nearer to the destination $u$.
Iterating this process, $u$ will be eventually reached and it will return the CIDs associated to $K$, in this case the reference to the Wikipedia page of the city of Rome stored in IPFS.
This type of punctual query is called Pin Search in \cite{joung2007keyword}, i.e. obtaining all and only the objects that are exactly associated with the keywords set $K$, $\{o \in O \mid K_o = K\}$; in this case data objects are retrieved only from one node. 

Another query type is the Superset Search. It is similar to a Pin Search, but in addition it also searches for objects that can be described by keywords sets that include $K$, i.e., $\{o \in O \mid K_o \supseteq K\}$. In this case, data objects are retrieved from all nodes that are responsible for a superset of $K$. Then, since the possible outcomes of this search can be quite large, a limit $l$ on the results is set. 
For instance, a Superset Search using the keywords set of the previous example would include:
\begin{itemize} 
    \item the objects retrieved from the node $u$ through the Pin Search 
    \item plus the objects retrieved from $u$'s neighbors, responsible for keywords sets such as $K_1=\{$\textit{``Wikipedia, Rome, PoI''}$\}$
    \item plus the objects retrieved from $u$'s neighbors' neighbors, with keywords sets such as $K_2=\{$\textit{``Wikipedia, Rome, PoI,Temperature''}$\}$
    \item and so on, until the number of objects is equal to $l$ or no more nodes shall be contacted.
\end{itemize} 

\section{Decentralized Autonomous Organization Framework}\label{sec:dao}
\begin{figure}
    \centering
    \includegraphics[width=.43\textwidth]{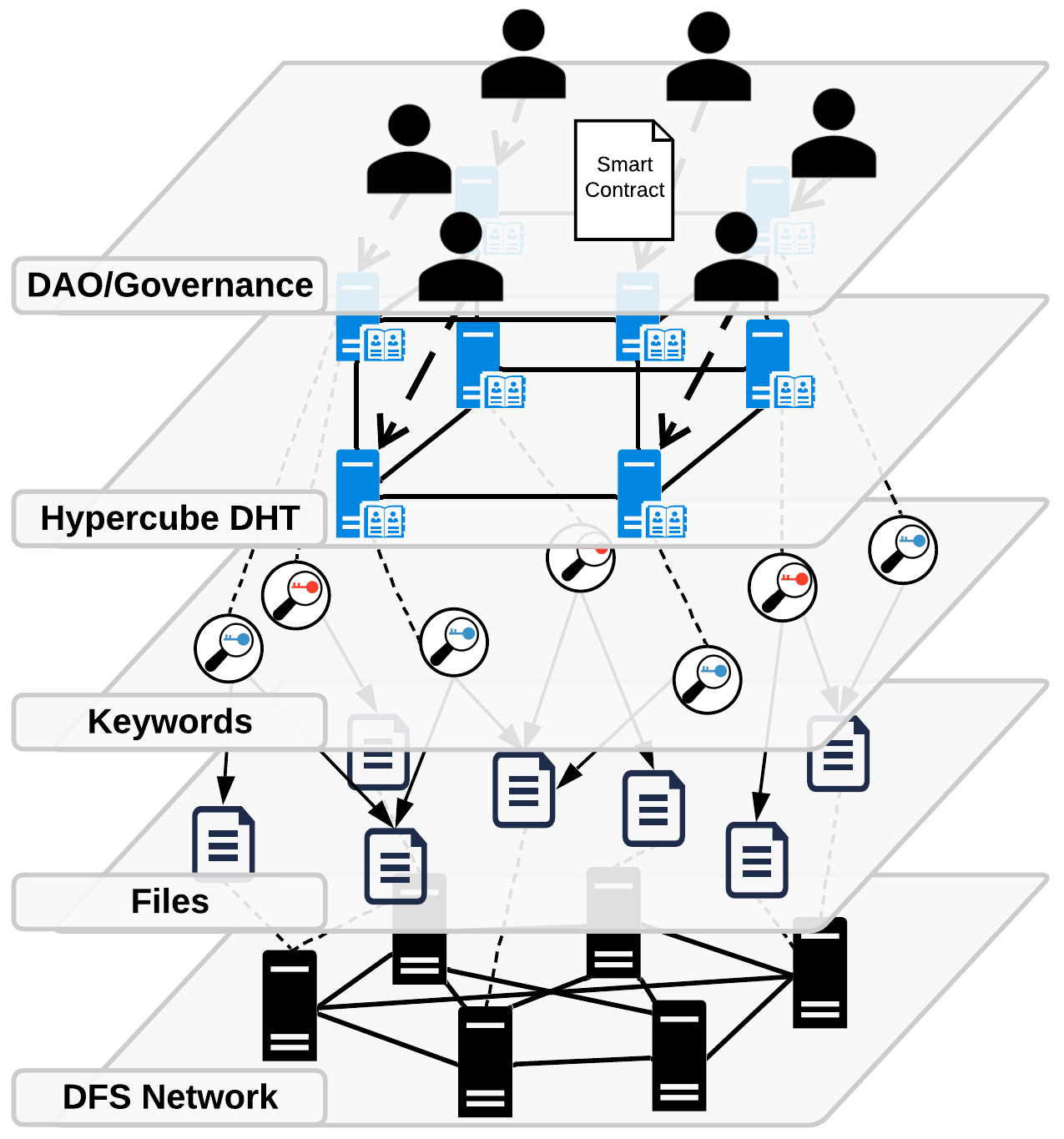}
    \caption{System represented as a layered architecture.}
    \label{fig:layers}
\end{figure}

The contribution presented so far in this paper can be represented as in Figure \ref{fig:layers} with the first four layers (bottom-up). To recap:
\begin{enumerate}
    \item First layer: the nodes of the IPFS public network running the standard IPFS protocol.
    \item Second layer: all the files that the IPFS nodes keep in their storage. These are indexed using CIDs.
    \item Third layer: the files can be described by keywords, which are then used to execute queries and find the files.
    \item Fourth layer: these keywords are saved, together with the file association via the CID, using the Hypercube DHT. 
\end{enumerate}
Before continuing, it is important to point out that this structure is independent from the IPFS implementation, but can be adapted to any type of DFS or DLT for data storage, e.g. IOTA DLT \cite{zichichi2020framework}.

The main focus of this section is the fifth layer. It consists of technologies and processes that form the governance of the hypercube DHT network, i.e. the DAO. Layers 1 to 4 provide the technological means for implementing a keyword based search over a DFS and this can be enough for offering a complete solution in many cases. However, we are also interested in the scenario where, in order to orchestrate the operational decisions and rewards, the DHT network nodes operators can form a DAO. The fifth layer is mostly based on the use of smart contracts and the interfaces to those. Smart contracts, indeed, enable the creation of an organization that takes advantage of a token-based economy and decentralized voting.
In particular, we use Ethereum smart contracts and we structure the DAO on top of two preliminary research contributions \cite{zichichi2019likestarter,distefano2020moatcoin}:
\begin{itemize}
    \item \textbf{Token economy} - The DAO is built around the use of a unique token, e.g. ``DAOToken'', used for transferring value (e.g.~users that pay node operators), or for staking purposes (e.g.~becoming a DAO member by time-locking a certain amount of tokens). The smart contract used to represent these functions consists in an implementation of the ERC20 interface \cite{erc202015}.
    \item \textbf{Members Registry} - A smart contract was developed as a members registry, to allow token holders to time-lock DAOTokens and become DAO members. Any account holding any amount of DAOToken can lock some tokens for a desired amount of time through a specific time-lock contract. This time-lock contract will hold these tokens and release them after the date set, and no one will be able to unlock those before that date.
    \item \textbf{General Voting} - A specific smart contract was developed to allow DAO members to call for a vote and then decide on a proposal. This contract allows any member to make a proposal and gives everyone the opportunity to submit a suggestion to vote regarding that proposal. Each proposal has its own debate period and any member can vote a suggestion within that time period. A member vote weight is proportional to the amount of tokens locked until a date that comes after the debate period end.
    \item \textbf{Value Transfer Voting} - Any extension of the previous voting smart contract can be developed to allow a decision taken to directly enact an operation to be executed on-chain (through another smart contract). For instance, DAO members can vote to transfer some staked tokens to a specific account in the case of issuing a bounty.
\end{itemize}

\subsection{Use Cases}
In this section we discuss on different, but possibly overlapping, use cases for implementing the DAO framework presented above. For instance, the first use case on DeFi is generally applicable to different DAOs implementations, and can be combined with the second use case to create a complete Hypercube DAO system.

\subsubsection{A DeFi-based rewarding system}
Decentralized finance (DeFi) is a term that refers to novel P2P financial infrastructures, based on smart contracts, that are non-custodial, permissionless, openly verifiable and composable \cite{werner2021sok}. With DeFi protocols such as Decentralized Exchanges (DEX), anyone can engage in non-custodial exchange of on-chain digital assets, e.g. tokens.
In contrast to traditional finance where an asset's liquidity is based on the bid and ask orders prices, in the most used DeFi protocols, such as Uniswap, usually assets are ERC20 tokens and their liquidity is provided algorithmically through a simple pricing rule within a smart contract \cite{werner2021sok}.
For instance, in the case of the DAOToken, an Uniswap liquidity pool smart contract can be created by locking into it an amount $x$ of DAOTokens and an amount $y$ of another ERC20 token to be exchanged with. The value of a single DAOToken in respect to the other token will be proportional to the ratio $\frac{y}{x}$ and such values will vary based on the tokens that will be stacked in the pool after an exchange, e.g. buying DAOTokens will drain the reserve of the locked DAOTokens and increase the other token's reserve.

In Uniswap, each liquidity provider receives newly minted Liquidity Pool (LP) tokens to represent the share of liquidity they have provided. These LP tokens can then be burn by the providers in order to redeem their share of liquidity (and accrued fees obtained when exchanges happen). This means that when a new ERC20 token, e.g. the DAOToken, is created and initially distributed to its creators, they can easily have a return on their newly created tokens by locking them in a new liquidity pool. This is also a way to let the general investors interested in this token to acquire it. However, the possibility for the creators to redeem at any time the liquidity they have provided, by burning the LP tokens, makes the value of the token highly unstable. At any moment, indeed, the investors can be left with a worthless token due to these ``big players'' burning LP tokens and draining the reserve.

Based on this, we envision a use case where the DAO is based not on the timelock of the DAOToken directly, but on the timelock of the LP tokens obtained by locking DAOTokens in liquidity pools. From an implementation point of view, in Uniswap no changes are required because LP Tokens are compatible with the ERC20 inteface. This means that the stability of the DAO is directly proportional to the value the DAOToken can take, and that the power exercised by DAO members is directly proportional to the gains/losses they are willing to make through their behavior, making it possible to have a strong incentive to behave correctly.

\subsubsection{Unique general purpose DAO vs. DAO islands}
The proposal we provide in this work comprises the use case where a unique DHT network is governed by a DAO, with the purpose of assisting ``browsable sister network[s] to the internet'' \cite{williams2018arweave}. Indeed, apart from IPFS, several DFS (and DLT-backed DFS such as Arweave), are built for the replication of the content that can be found on Internet. In this use case we propose a unique DAO that deals with the maintaining of a DHT that allows to search through keywords in those general purpose file storages.

Opposed to the previous one, we envision a use case where different networks implement their own DHT, resulting in a multitude of ``islands'' where keywords-based queries are possible for specific topics or platforms. Each island has its own organization and rules, but they are all similar for the querying protocol. It could be the case in which several smart contracts enabled DLTs are used for making the DAO, or, conversely, only one DLT used, with the same token shared among the different DAOs. An example would be a DAO maintaining the hypercube DHT for querying the decentralized version of Wikipedia, or a DAO for maintaining querying for political content shared in social media.

\begin{table}[t]
\centering
\begin{tabular}{ |c|c|c| } 
\hline
\textbf{Smart Contract} & \textbf{Operation} & \textbf{Cost (gas)} \\ 
\hline
ERC20 & transfer() &  51167 \\
\hline
TokenTimelockProxy & lockTokens() &  232024 \\
\hline
TokenTimelock & release() &  25626 \\
\hline
Voting & submitProposal() &  133501 \\
\hline
Voting & submitSuggestion() &  114523 \\
\hline
Voting & vote() &  142848 \\
\hline
Voting & executeProposal() &  56991 \\
\hline
\end{tabular}
\caption{DAO smart contracts operations cost in terms of gas.}
\label{table:gas}
\end{table}

\subsubsection{Decentralized ipfs-search}
Finally, a possible use case is based on the possibility of combining ipfs-search~\cite{ipfssearch2021} with our keyword-based hypercube DHT.
That is, we consider a protocol where:
\begin{itemize}
    \item several IPFS nodes crawl the network by monitoring the IPFS logs for files addition;
    \item these nodes download the files added through the ``sniffed'' CID;
    \item for each file the metadata are extracted and transformed into keywords;
    \item the association between the keywords obtained and the CID is stored in the hypercube DHT.
\end{itemize}

\section{Experimental Evaluation}\label{sec:valid}

\begin{figure*}[ht!]
    \centering
    \includegraphics[width=.9\textwidth]{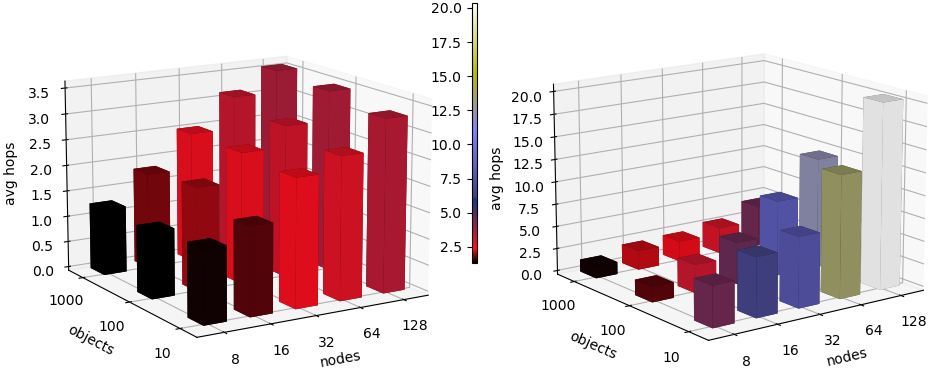}
    \caption{Pin Search results on the left, Superset Search on the right.}
    \label{fig:plot}
\end{figure*}

In this section, we provide an experimental validation of our work. In particular, we implemented the software that each hypercube DHT logical node runs for maintaining the index table and to answer the queries that it receives. Furthermore, we developed a prototype of the DAO framework.

\subsection{DAO Smart Contracts}
The smart contracts that implement the framework presented above have been developed in Solidity and stored as Open Source code in Zenodo \cite{mirko_zichichi_2021_4767755}. In Table \ref{table:gas}, we provide the cost execution in terms of gas \cite{buterin2013ethereum} for the main operations. The most expensive operation is the \textit{lockTokens()} function, that locks a certain amount of an ERC20 Token for a specified amount of time. This is because, following the OpenZeppelin library for secure smart contracts development \cite{openzeppelin2021}, each lock request creates a new smart contract that locks tokens for a unique account. However, normally the creation and deployment of such a smart contract through the Factory pattern would require at least $501818$ gas units. We used the EIP-1167 Minimal Proxy pattern \cite{eip11672018}, that, instead of deploying a new contract each time such as in the Factory pattern, clones an already deployed contract functionalities by delegating all function calls to it.

\subsection{Hypercube}
The DHT software is implemented in Python and it exposes the four main nodes actions using the Flask server framework \cite{grinberg2018flask}, i.e. Insert object, Remove object, Pin search, Superset search. Together with the core logic methods for a logical node, the implementation includes also an interface for communicating with an IPFS node, in order to possibly return files instead of only CIDs. In the same physical nodes, it is possible to host more than one logical node, that might refer to the same (local) IPFS node.

We tested our implementation running the software on a dedicated host (i.e.~a quad core CPU, 16GB RAM), by associating several logical nodes to different operative system ports. Each logical node was executed by a dedicated Flask server and after a bootstrap phase, each node was connected to its neighbors based on the hypercube topology. More specifically, we run two different types of tests, one for the Pin Search and one for the Superset Search. In both cases, we tested the network configuration for $8$, $16$, $32$, $64$ and $128$ nodes and populated the network each time with $10$, $100$ and $1000$ objects generated randomly. Then, we repeated 50 random queries for Pin Search and 50 for Superset Search (with objects limit set to 10). During each query, a node was randomly chosen and it was queried using a random keywords set.

The results for the test that we carried out are reported in Figure \ref{fig:plot}. These results are obtained averaging the number of hops needed to get from on node in the network to another, for different operations. 

\subsubsection{Pin Search}
Results for the Pin Search (Figure \ref{fig:plot}, left) show similar average hops when the number of objects varies and an increase from $1.28$ to $3.52$ when increasing the number of nodes. This was an expected result as the Pin Search average number of hops should theoretically be with the order of the logarithm of the number of logical nodes, i.e. $\frac{\log(n)}{2}$ or $\frac{r}{2}$. For instance, with $128$ nodes, the experienced average number of hops was around $\frac{\log(128)}{2} = 3.5$.

\subsubsection{Superset Search}
In the case of Superset Search results are different from the previous case. As Figure \ref{fig:plot} (right) shows, the average number of hops decreases when the number of objects increases, and it increases when the number of nodes increases. The minimum value here is $1.36$ for $1000$ objects and $8$ nodes and the maximum is $20.36$ for $10$ objects and $128$ nodes. Theoretically, the average number of hops should be equal to the average hops required to get to the node responsible for query keywords set $K$, i.e. Pin Search $\frac{\log(n)}{2}$, plus the average hops to get from that node to all the nodes that that include $K$, until the limit of objects $l$ (or nodes including $K$) is reached.

\section{Conclusions}\label{sec:concl}
In this work, we proposed a decentralized system that manages keyword-based queries for contents stored in IPFS, through the use of an hypercube DHT. 
The query routing efficiency lies in the traversal of the hypercube which has a maximum number of hops of $\log(\textit{number of nodes}) = r$, i.e. the hypercube dimension.
Our experimental validation is in line with this number and shows that on average $\frac{r}{2}$ hops are required for the Pin Search. While, in the case of the Superset Search, we experienced the dependence of the number of hops with the ratio between the limit $l$ assigned to the query and the distribution of objects between nodes.

Furthermore, we described the development of a DAO related to the economic sustainability and development of the project, as well as use cases for the government of the above system. The use of Ethereum smart contracts enables the possibility of voting for making organizational decisions. Furthermore, the ability to create ERC20 tokens allows to reward nodes that have actively contributed to the operation of the P2P system.

As a future work, we will focus on two aspects. Firstly, we will investigate on the feasibility of a ``pay-per-query'' model, where node operators within the DAO are rewarded at the level of granularity of the query. Secondly, we will face load balancing issues that arise when a more realistic contents distribution is put in place and where some nodes might suffer an higher workload due to the popularity of the contents/keywords they store.

\bibliographystyle{ACM-Reference-Format}
\bibliography{sample-base}


\begin{thebibliography}{23}


\ifx \showCODEN    \undefined \def \showCODEN     #1{\unskip}     \fi
\ifx \showDOI      \undefined \def \showDOI       #1{#1}\fi
\ifx \showISBNx    \undefined \def \showISBNx     #1{\unskip}     \fi
\ifx \showISBNxiii \undefined \def \showISBNxiii  #1{\unskip}     \fi
\ifx \showISSN     \undefined \def \showISSN      #1{\unskip}     \fi
\ifx \showLCCN     \undefined \def \showLCCN      #1{\unskip}     \fi
\ifx \shownote     \undefined \def \shownote      #1{#1}          \fi
\ifx \showarticletitle \undefined \def \showarticletitle #1{#1}   \fi
\ifx \showURL      \undefined \def \showURL       {\relax}        \fi
\providecommand\bibfield[2]{#2}
\providecommand\bibinfo[2]{#2}
\providecommand\natexlab[1]{#1}
\providecommand\showeprint[2][]{arXiv:#2}

\bibitem[\protect\citeauthoryear{Belotti, Bo{\v{z}}i{\'c}, Pujolle, and
  Secci}{Belotti et~al\mbox{.}}{2019}]%
        {belotti2019vademecum}
\bibfield{author}{\bibinfo{person}{Marianna Belotti}, \bibinfo{person}{Nikola
  Bo{\v{z}}i{\'c}}, \bibinfo{person}{Guy Pujolle}, {and}
  \bibinfo{person}{Stefano Secci}.} \bibinfo{year}{2019}\natexlab{}.
\newblock \showarticletitle{A vademecum on blockchain technologies: When,
  which, and how}.
\newblock \bibinfo{journal}{\emph{IEEE Communications Surveys \& Tutorials}}
  \bibinfo{volume}{21}, \bibinfo{number}{4} (\bibinfo{year}{2019}),
  \bibinfo{pages}{3796--3838}.
\newblock


\bibitem[\protect\citeauthoryear{Benet}{Benet}{2014}]%
        {benet2014ipfs}
\bibfield{author}{\bibinfo{person}{Juan Benet}.}
  \bibinfo{year}{2014}\natexlab{}.
\newblock \showarticletitle{Ipfs-content addressed, versioned, p2p file
  system}.
\newblock \bibinfo{journal}{\emph{arXiv preprint arXiv:1407.3561}}
  (\bibinfo{year}{2014}).
\newblock


\bibitem[\protect\citeauthoryear{Buterin}{Buterin}{2013}]%
        {buterin2013ethereum}
\bibfield{author}{\bibinfo{person}{Vitalik et~al. Buterin}.}
  \bibinfo{year}{2013}\natexlab{}.
\newblock \bibinfo{title}{Ethereum white paper}.
\newblock
\newblock
\urldef\tempurl%
\url{https://github.com/ethereum/wiki/wiki/White-Paper}
\showURL{%
\tempurl}


\bibitem[\protect\citeauthoryear{Distefano, Pocher, and Zichichi}{Distefano
  et~al\mbox{.}}{2020}]%
        {distefano2020moatcoin}
\bibfield{author}{\bibinfo{person}{Biagio Distefano}, \bibinfo{person}{Nadia
  Pocher}, {and} \bibinfo{person}{Mirko Zichichi}.}
  \bibinfo{year}{2020}\natexlab{}.
\newblock \showarticletitle{{MOATcoin: Exploring Challenges and Legal
  Implications of Smart Contracts Through a Gamelike DApp Experiment}}. In
  \bibinfo{booktitle}{\emph{Proc. of the 3rd Workshop on Cryptocurrencies and
  Blockchains for Distributed Systems (CryBlock 2020), co-located with the 26th
  Annual International Conference on Mobile Computing and Networking (MobiCom
  2020), ACM}}. ACM, \bibinfo{pages}{1--6}.
\newblock


\bibitem[\protect\citeauthoryear{D’Angelo and Ferretti}{D’Angelo and
  Ferretti}{2017}]%
        {d2017highly}
\bibfield{author}{\bibinfo{person}{Gabriele D’Angelo} {and}
  \bibinfo{person}{Stefano Ferretti}.} \bibinfo{year}{2017}\natexlab{}.
\newblock \showarticletitle{Highly intensive data dissemination in complex
  networks}.
\newblock \bibinfo{journal}{\emph{J. Parallel and Distrib. Comput.}}
  \bibinfo{volume}{99} (\bibinfo{year}{2017}), \bibinfo{pages}{28--50}.
\newblock


\bibitem[\protect\citeauthoryear{Fabian~Vogelsteller}{Fabian~Vogelsteller}{2015}]%
        {erc202015}
\bibfield{author}{\bibinfo{person}{Vitalik~Buterin Fabian~Vogelsteller}.}
  \bibinfo{year}{2015}\natexlab{}.
\newblock \bibinfo{title}{EIP-20: ERC-20 Token Standard}.
\newblock
\newblock
\urldef\tempurl%
\url{https://eips.ethereum.org/EIPS/eip-20}
\showURL{%
\tempurl}


\bibitem[\protect\citeauthoryear{Grinberg}{Grinberg}{2018}]%
        {grinberg2018flask}
\bibfield{author}{\bibinfo{person}{Miguel Grinberg}.}
  \bibinfo{year}{2018}\natexlab{}.
\newblock \bibinfo{booktitle}{\emph{Flask web development: developing web
  applications with python}}.
\newblock \bibinfo{publisher}{" O'Reilly Media, Inc."}.
\newblock


\bibitem[\protect\citeauthoryear{Guidi, Michienzi, and Ricci}{Guidi
  et~al\mbox{.}}{2021}]%
        {guidi2021data}
\bibfield{author}{\bibinfo{person}{Barbara Guidi}, \bibinfo{person}{Andrea
  Michienzi}, {and} \bibinfo{person}{Laura Ricci}.}
  \bibinfo{year}{2021}\natexlab{}.
\newblock \showarticletitle{Data Persistence in Decentralized Social
  Applications: The IPFS approach}. In \bibinfo{booktitle}{\emph{2021 IEEE 18th
  Annual Consumer Communications \& Networking Conference (CCNC)}}. IEEE,
  \bibinfo{pages}{1--4}.
\newblock


\bibitem[\protect\citeauthoryear{Hamari, Sj{\"o}klint, and Ukkonen}{Hamari
  et~al\mbox{.}}{2016}]%
        {hamari2016sharing}
\bibfield{author}{\bibinfo{person}{Juho Hamari}, \bibinfo{person}{Mimmi
  Sj{\"o}klint}, {and} \bibinfo{person}{Antti Ukkonen}.}
  \bibinfo{year}{2016}\natexlab{}.
\newblock \showarticletitle{The sharing economy: Why people participate in
  collaborative consumption}.
\newblock \bibinfo{journal}{\emph{Journal of the association for information
  science and technology}} \bibinfo{volume}{67}, \bibinfo{number}{9}
  (\bibinfo{year}{2016}), \bibinfo{pages}{2047--2059}.
\newblock


\bibitem[\protect\citeauthoryear{{IPFS Community}}{{IPFS Community}}{2021}]%
        {ipfssearch2021}
\bibfield{author}{\bibinfo{person}{{IPFS Community}}.}
  \bibinfo{year}{2021}\natexlab{}.
\newblock \bibinfo{title}{Search engine for the InterPlanetary File System}.
\newblock
  \bibinfo{howpublished}{\url{https://github.com/ipfs-search/ipfs-search}}.
\newblock


\bibitem[\protect\citeauthoryear{Jentzsch}{Jentzsch}{2016}]%
        {jentzsch2016decentralized}
\bibfield{author}{\bibinfo{person}{Christoph Jentzsch}.}
  \bibinfo{year}{2016}\natexlab{}.
\newblock \showarticletitle{Decentralized autonomous organization to automate
  governance}.
\newblock \bibinfo{journal}{\emph{White paper, November}}
  (\bibinfo{year}{2016}).
\newblock
\urldef\tempurl%
\url{https://lawofthelevel.lexblogplatformthree.com/wp-content/uploads/sites/187/2017/07/WhitePaper-1.pdf}
\showURL{%
\tempurl}


\bibitem[\protect\citeauthoryear{Joung, Yang, and Fang}{Joung
  et~al\mbox{.}}{2007}]%
        {joung2007keyword}
\bibfield{author}{\bibinfo{person}{Yuh-Jzer Joung}, \bibinfo{person}{Li-Wei
  Yang}, {and} \bibinfo{person}{Chien-Tse Fang}.}
  \bibinfo{year}{2007}\natexlab{}.
\newblock \showarticletitle{Keyword search in dht-based peer-to-peer networks}.
\newblock \bibinfo{journal}{\emph{IEEE Journal on Selected Areas in
  Communications}} \bibinfo{volume}{25}, \bibinfo{number}{1}
  (\bibinfo{year}{2007}), \bibinfo{pages}{46--61}.
\newblock


\bibitem[\protect\citeauthoryear{Khudhur and Fujita}{Khudhur and
  Fujita}{2019}]%
        {khudhur2019siva}
\bibfield{author}{\bibinfo{person}{Nawras Khudhur} {and}
  \bibinfo{person}{Satoshi Fujita}.} \bibinfo{year}{2019}\natexlab{}.
\newblock \showarticletitle{Siva-The IPFS Search Engine}. In
  \bibinfo{booktitle}{\emph{2019 Seventh International Symposium on Computing
  and Networking (CANDAR)}}. IEEE, \bibinfo{pages}{150--156}.
\newblock


\bibitem[\protect\citeauthoryear{OpenZeppelin}{OpenZeppelin}{2021}]%
        {openzeppelin2021}
\bibfield{author}{\bibinfo{person}{OpenZeppelin}.}
  \bibinfo{year}{2021}\natexlab{}.
\newblock \bibinfo{title}{OpenZeppelin website}.
\newblock
\newblock
\urldef\tempurl%
\url{https://openzeppelin.com/}
\showURL{%
\tempurl}


\bibitem[\protect\citeauthoryear{Pazaitis, De~Filippi, and Kostakis}{Pazaitis
  et~al\mbox{.}}{2017}]%
        {pazaitis2017blockchain}
\bibfield{author}{\bibinfo{person}{Alex Pazaitis}, \bibinfo{person}{Primavera
  De~Filippi}, {and} \bibinfo{person}{Vasilis Kostakis}.}
  \bibinfo{year}{2017}\natexlab{}.
\newblock \showarticletitle{Blockchain and value systems in the sharing
  economy: The illustrative case of Backfeed}.
\newblock \bibinfo{journal}{\emph{Technological Forecasting and Social Change}}
   \bibinfo{volume}{125} (\bibinfo{year}{2017}), \bibinfo{pages}{105--115}.
\newblock


\bibitem[\protect\citeauthoryear{Peter~Murray}{Peter~Murray}{2018}]%
        {eip11672018}
\bibfield{author}{\bibinfo{person}{Joe~Messerman Peter~Murray, Nate~Welch}.}
  \bibinfo{year}{2018}\natexlab{}.
\newblock \bibinfo{title}{EIP-1167: Minimal Proxy Contract}.
\newblock
\newblock
\urldef\tempurl%
\url{https://eips.ethereum.org/EIPS/eip-1167}
\showURL{%
\tempurl}


\bibitem[\protect\citeauthoryear{Santos, Santos, and Dias}{Santos
  et~al\mbox{.}}{2019}]%
        {santos2019dclaims}
\bibfield{author}{\bibinfo{person}{Jo{\~a}o Santos}, \bibinfo{person}{Nuno
  Santos}, {and} \bibinfo{person}{David Dias}.}
  \bibinfo{year}{2019}\natexlab{}.
\newblock \showarticletitle{DClaims: A censorship resistant web annotations
  system using IPFS and Ethereum}.
\newblock \bibinfo{journal}{\emph{arXiv preprint arXiv:1912.03388}}
  (\bibinfo{year}{2019}).
\newblock


\bibitem[\protect\citeauthoryear{{The Graph}}{{The Graph}}{2020}]%
        {thegraph2020protocol}
\bibfield{author}{\bibinfo{person}{{The Graph}}.}
  \bibinfo{year}{2020}\natexlab{}.
\newblock \bibinfo{title}{The Graph Protocol}.
\newblock
\newblock
\urldef\tempurl%
\url{https://thegraph.com/}
\showURL{%
\tempurl}


\bibitem[\protect\citeauthoryear{Werner, Perez, Gudgeon, Klages-Mundt, Harz,
  and Knottenbelt}{Werner et~al\mbox{.}}{2021}]%
        {werner2021sok}
\bibfield{author}{\bibinfo{person}{Sam~M Werner}, \bibinfo{person}{Daniel
  Perez}, \bibinfo{person}{Lewis Gudgeon}, \bibinfo{person}{Ariah
  Klages-Mundt}, \bibinfo{person}{Dominik Harz}, {and}
  \bibinfo{person}{William~J Knottenbelt}.} \bibinfo{year}{2021}\natexlab{}.
\newblock \showarticletitle{SoK: Decentralized Finance (DeFi)}.
\newblock \bibinfo{journal}{\emph{arXiv preprint arXiv:2101.08778}}
  (\bibinfo{year}{2021}).
\newblock


\bibitem[\protect\citeauthoryear{Williams and Jones}{Williams and
  Jones}{2018}]%
        {williams2018arweave}
\bibfield{author}{\bibinfo{person}{Samuel Williams} {and}
  \bibinfo{person}{William Jones}.} \bibinfo{year}{2018}\natexlab{}.
\newblock \showarticletitle{Arweave Lightpaper}.
\newblock  (\bibinfo{year}{2018}).
\newblock


\bibitem[\protect\citeauthoryear{Zichichi}{Zichichi}{2021}]%
        {mirko_zichichi_2021_4767755}
\bibfield{author}{\bibinfo{person}{Mirko Zichichi}.}
  \bibinfo{year}{2021}\natexlab{}.
\newblock \bibinfo{booktitle}{\emph{miker83z/HypercubeDAOContracts}}.
\newblock
\urldef\tempurl%
\url{https://doi.org/10.5281/zenodo.4767755}
\showDOI{\tempurl}


\bibitem[\protect\citeauthoryear{Zichichi, Contu, Ferretti, and
  D'Angelo}{Zichichi et~al\mbox{.}}{2019}]%
        {zichichi2019likestarter}
\bibfield{author}{\bibinfo{person}{Mirko Zichichi}, \bibinfo{person}{Michele
  Contu}, \bibinfo{person}{Stefano Ferretti}, {and} \bibinfo{person}{Gabriele
  D'Angelo}.} \bibinfo{year}{2019}\natexlab{}.
\newblock \showarticletitle{LikeStarter: a {Smart-contract} based Social {DAO}
  for Crowdfunding}. In \bibinfo{booktitle}{\emph{Proc. of the 2st Workshop on
  Cryptocurrencies and Blockchains for Distributed Systems}}.
\newblock


\bibitem[\protect\citeauthoryear{Zichichi, Ferretti, and D’Angelo}{Zichichi
  et~al\mbox{.}}{2020}]%
        {zichichi2020framework}
\bibfield{author}{\bibinfo{person}{Mirko Zichichi}, \bibinfo{person}{Stefano
  Ferretti}, {and} \bibinfo{person}{Gabriele D’Angelo}.}
  \bibinfo{year}{2020}\natexlab{}.
\newblock \showarticletitle{A Framework based on Distributed Ledger
  Technologies for Data Management and Services in Intelligent Transportation
  Systems}.
\newblock \bibinfo{journal}{\emph{IEEE Access}} (\bibinfo{year}{2020}).
\newblock


\end{thebibliography}

\end{document}